\def\o{\over}
\def\A{\rightarrow}
\def\bar{\overline}

\def\r{\gamma}
\def\a{\alpha}
\def\b{\beta}

\def\th{\theta}
\def\vp{\varphi}
\def\Re{{\rm Re}}
\def\Im{{\rm Im}}

\def\bar{\overline}

\def\G{{\rm GeV}}
\documentstyle [12pt]{article}
\begin{document}
\baselineskip=25pt
\setcounter{page}{1}
\thispagestyle{empty}
\topskip 0.0  cm
\begin{flushright}
\begin{tabular}{c c}
& {\normalsize   UWThPh-1994-25}\\
& {\normalsize   AUE-04-94}\\
& \today
\end{tabular}
\end{flushright}
\vspace{-0.2cm}
\centerline{\Large\bf Maximal $CP$ violation in the Higgs Sector}
\centerline{\Large\bf and its Effect on the $\rho$ Parameter}
\vskip 0.1 cm
\centerline{{\bf Girish C. JOSHI}$^{(a)}$
\footnote{E-mail:joshi@bradman.phys.unimelb..ac.au}
{\bf Masahisa  MATSUDA}$^{(b)}$
              \footnote{E-mail:masa@auephyas.aichi-edu.ac.jp}}
\centerline{and}
\centerline{{\bf Morimitsu TANIMOTO}$^{(c)}$
  \footnote{Permanent address:Science Education Laboratory, Ehime University,
790 Matsuyama, JAPAN}}
 \vskip 0.1 cm
\centerline{$^{(a)}$ \it{Department of Physics, University of Melbourne}}
\centerline{\it Parkville, Victoria 3052, AUSTRALIA}
\centerline{$^{(b)}$ \it{Department of Physics and
      Astronomy, Aichi University of Education}}
\centerline{\it Kariya, Aichi 448, JAPAN}
\centerline{$^{(c)}$ \it{Institut f\"ur Theoretische Physik,
               Universit\"at Wien}}
\centerline{ \it Boltzmanngasse 5, A-1090 Wien, AUSTRIA}

\vskip 0.3 cm
\centerline{\bf ABSTRACT}
We  study the conditions of maximal $CP$ violation in the neutral Higgs mass
matrix of the two Higgs doublet model. We get fixed values of $\tan\b$  and
  constraints on the Higgs potential parameters.
 Two  neutral Higgs scalars are constrained to be lighter  than  the charged
Higgs scalar and these two Higgs scalars
 are expected to be
almost degenerate due to the smallness of the $h$ parameter, where $h$ is the
CP violating coupling constant of the Higgs interaction.
The radiative correction of the $\rho$ parameter from the Higgs scalar
exchange  is rather small and its sign negative for a wide
range of Higgs masses.
It follows that maximum $CP$ violation in the two Higgs
doublet model is safely
allowed for the $\rho$ parameter without the custodial symmetry.
\newpage
\topskip 1 cm
  The physics of  $CP$ violation has attracted much recent attention
in the light that the $B$-factory will go on line in the near future.
In the standard model(SM), the origin of such $CP$
violation is reduced to the phase in the Kobayashi-Maskawa matrix[1]. However,
 there has been a general interest in considering other approaches to $CP$
violation since many alternate sources exist.\par
  In this paper, we consider the simplest and most attractive extension of the
 standard Higgs sector, namely the type II two Higgs doublet model(THDM)[2],
yielding both charged and neutral Higgs bosons as physical states.
The THDM with the soft breaking term of the discrete symmetry demonstrates
explicit or spontaneous $CP$ violation[3].
 Some authors have proposed to search for the $CP$-violating observables
in the Higgs sectors, where physical effects occur in the
electric dipole moment of the neutron[4] and
electron[5], or as asymmetries in the top quark production or decay[6].
Such loop contributions are generally small and decrease with increasing
Higgs masses.
Direct $CP$-violating Higgs productions were also predicted
in $e^+e^-$ colliders[7,8]. On the other hand, the radiative corrections
arising from the Higgs scalar contribution to the vector-boson self-enegyies
have been studied in this model[9].  In this context, Pomarol and Vega proposed
   a new way to constrain $CP$ violation in the Higgs sector using
the experimental value of the $\rho$
   parameter[10]. Since a custodial $SU(2)$ symmetry cannot be defined in the
$CP$-violating Higgs potential, the radiative corrections to $\rho$ are
unavoidable.
\par
 In this paper, we study the neutral Higgs mass matrix with maximal
$CP$ violation and its effect on  the $\rho$ parameter
 in the THDM[7,10].
We have found that maximal $CP$ violation is realized
 under the fixed values of  $\tan\b$ with two  constraints of parameters in the
   Higgs potential. Taking these conditions into account, we
investigate the contribution of the Higgs sector to the $\rho$ parameter. \par
First, we will discuss maximal $CP$ violation in the THDM.
The Higgs potential with $CP$ violating terms in the THDM can be written as:
\begin{eqnarray}
V _{{\rm Higgs}}&=&{1\o 2}g_1(\Phi_1^\dagger\Phi_1-|v_1|^2)^2+
         {1\o 2}g_2(\Phi_2^\dagger\Phi_2-|v_2|^2)^2 \nonumber\\
  &+& g(\Phi_1^\dagger\Phi_1-|v_1|^2)(\Phi_2^\dagger\Phi_2-|v_2|^2)
      \nonumber\\
&+& g'|\Phi_1^\dagger\Phi_2-v_1^*v_2|^2+Re[h(\Phi_1^\dagger\Phi_2-v_1^*v_2)^2]
           \nonumber\\
  &+& \xi\left [{\Phi_1\o v_1}-{\Phi_2\o v_2}\right ]^\dagger
         \left [{\Phi_1\o v_1}-{\Phi_2\o v_2}\right ] \ ,
\end{eqnarray}
\noindent where
$\Phi_1$ and $\Phi_2$
couple with the down-quark and the up-quark sectors respectively
and   the vacuum expectation values are defined as
$v_1\equiv <\Phi_1^0>_{vac}$  and  $v_2\equiv <\Phi_2^0>_{vac}$.
 We do not concern ourselves here with a specific model of $CP$
violation, but instead consider a general parametrization using the
notation developed by Weinberg[11].
We take $h$ to be real and set
\begin{equation}
  v_1^* v_2=|v_1 v_2|\exp(i\phi) \ ,
\end{equation}
\noindent as a phase convension.
We define the neutral components of the two Higgs doublets
using three real fields $\phi_1,\phi_2,\phi_3$
 and the Goldstone boson $\chi^0$ as follows:
 \begin{eqnarray}
\Phi_1^0&=&{1\o\sqrt{2}}\{\phi_1+\sqrt{2}v_1+i(\cos\b\chi^0-\sin\b\phi_3)\}\ ,
        \nonumber\\
\Phi_2^0&=&{1\o\sqrt{2}}\{\phi_2+\sqrt{2}v_2+i(\sin\b\chi^0+\cos\b\phi_3)\}\ ,
\end{eqnarray}
\noindent
 where $\tan\b\equiv v_2/v_1$.
The real fields $\phi_1$ and $\phi_2$ are scalar particles while
$\phi_3$ is pseudo-scalar in the limit of $CP$ conservation.
$CP$ violation will occur via the scalar-pseudoscalar interference terms in
   the neutral Higgs mass matrix.
Maximal CP violation was  defined on a new basis by Georgi[12],
where now the Goldstone boson decouples from the
 $\Phi_2$ doublet.
The neutral Higgs scalars
$H^0, H^1, H^2$ on this new basis are given by the following
 rotation;
\begin{equation}
  \left( \matrix{H^0 \cr H^1\cr H^2} \right )
=\left(\matrix{\cos\b& \sin\b &0\cr -\sin\b& \cos\b &0\cr 0&0&1 }  \right)
\left
   ( \matrix{\phi_1\cr \phi_2\cr \phi_3} \right ) \ .
\end{equation}
Denoting the orthogonal matrix ${\bf O}$ that relates this basis
with the mass eigenstates(physical states)
$\vp_1$, $\vp_2$ and $\vp_3$ as
\begin{equation}
  \left( \matrix{H^0 \cr H^1\cr H^2} \right )
={\bf O} \left( \matrix{\vp_1\cr \vp_2\cr \vp_3} \right ) \ ,
\end{equation}
\noindent
  maximal $CP$ violation is defined when
\begin{equation}
  {\bf O}_{11}^2={\bf O}_{12}^2={\bf O}_{13}^2={1\o 3} \ ,
\end{equation}
\noindent
which was discussed by  M\'endez and Pomarol[7].
\par
In order to get maximal CP violation we investigate the
eigenvectors and eigenvalues, which are given by solving the $3\times 3$
neutral Higgs mass matrix.
The Higgs mass matrix elements of the neutral
Higgs mass matrix ${\bf M^2}$ in the  $\phi_1$, $\phi_2 $ and $\phi_3$ basis of
   eq.(3) are given by
\begin{eqnarray}
M_{11}^2&=&2g_1|v_1|^2+g'|v_2|^2+{\xi+\Re(hv_1^{*2}v_2^2)\o|v_1|^2}\ ,
        \nonumber\\
M_{22}^2&=&2g_2|v_2|^2+g'|v_1|^2+{\xi+\Re(hv_1^{*2}v_2^2)\o|v_2|^2}\ ,
        \nonumber\\
M_{33}^2&=&(|v_1|^2+|v_2|^2) \left [g'+
              {\xi-\Re(hv_1^{*2}v_2^2)\o|v_1v_2|^2}\right ]\ ,\nonumber\\
M_{12}^2&=&|v_1v_2|(2g+g')+{\Re(hv_1^{*2}v_2^2)-\xi\o|v_1v_2|}\ , \\
M_{13}^2&=&-{\sqrt{|v_1|^2+|v_2|^2}\o|v_1^2v_2|}\Im(hv_1^{*2}v_2^2)\ ,
         \nonumber\\
M_{23}^2&=&-{\sqrt{|v_1|^2+|v_2|^2}\o|v_1v_2^2|}\Im(hv_1^{*2}v_2^2)\ ,\nonumber
\end{eqnarray}
\noindent which is the symmetric mass matrix.
  Denoting the orthogonal matrix ${\bf U}$ to diagonalize this mass matrix,
${\bf O}$, which is defined in eq.(5), is obtained by
\begin{equation}
  {\bf O}
=\left(\matrix{\cos\b& \sin\b &0\cr -\sin\b& \cos\b &0\cr 0&0&1 }  \right)
{\bf U}  \ .
\end{equation}
\noindent We then have

\begin{eqnarray}
{\bf O}_{11}&=&\cos\b{\bf U}_{11}+\sin\b {\bf U}_{21}\ ,
        \nonumber\\
{\bf O}_{12}&=&\cos\b{\bf U}_{12}+\sin\b {\bf U}_{22}\ ,
             \nonumber\\
{\bf O}_{13}&=&\cos\b{\bf U}_{13}+\sin\b {\bf U}_{23}\ .
\end{eqnarray}

Let us consider the condition
for maximal $CP$ violation in the neutral Higgs sector.
The trivial case that ${\bf O}_{11}={\bf O}_{12}=
{\bf O}_{13}$ is realized by ${\bf U}_{i1}={\bf U}_{i2}=
{\bf U}_{i3}(i=1,2)$.
However, these relations for the matrix
${\bf U}$ are forbidden by unitarity.
So, we consider instead the case that two matrix elements among
${\bf O}_{1i}(i=1,2,3)$ are equal without fixing the value of $\tan\b$.
Then  the value of $\tan\b$ can be tuned so as to give three
equal matrix elements.
First, we study
the case of  ${\bf O}_{12}={\bf O}_{13}$.
This relation is obtained if
\begin{equation}
  {\bf U}_{12}={\bf U}_{13} \ , \quad
  {\bf U}_{22}={\bf U}_{23} \quad
\end{equation}
\noindent or
\begin{equation}
\tan\b={{\bf U}_{12}-{\bf U}_{13}\o {\bf U}_{23}-{\bf U}_{22}}
  \ .
\end{equation}
\noindent
In the case of eq.(10), the orthogonal matrix ${\bf U}$ is specified as
follows:
\begin{equation}
 {\bf U}=\left(\matrix{\cos\th& {1\o \sqrt{2}}\sin\th &
   {1\o \sqrt{2}}\sin\th\cr -\sin\th& {1\o \sqrt{2}}\cos\th &
{1\o \sqrt{2}}\cos\th \cr 0&-{1\o \sqrt{2}} & {1\o \sqrt{2}} }
   \right)=
 \left(\matrix{\cos\th& \sin\th &0\cr -\sin\th& \cos\th &0\cr 0&0&1 } \right)
 \left(\matrix{1& 0 &0\cr 0&{1\o \sqrt{2}}  &
{1\o \sqrt{2}}\cr 0&-{1\o \sqrt{2}}&{1\o \sqrt{2}}} \right) \ ,
\end{equation}
\noindent
where $\th$ is an arbitrary rotation angle.
 It is easy to show  that this orthogonal matrix is obtained when
   both $M_{12}^2=0$ and $M_{13}^2=0$ are derived after rotating ${\bf M^2}$ on
the (1-2) plane with the angle $\th$.
The maximal mixing is then given on the new (2-3) plane.
We can easily find this case in the Higgs mass matrix
of eq.(7).\par
In the case of eq.(11), the situation is somewhat complicated.
 Let us denote by ${\bf V}$ the orthogonal matrix required to diagonalize
the Higgs
mass matrix on the new basis after rotating ${\bf M^2}$ on the (1-2) plane
with the angle $-\b$.
We then have
\begin{eqnarray}
{\bf U}_{12}-{\bf U}_{13}&=&\cos\b({\bf V}_{12}-{\bf V}_{13})
  - \sin\b({\bf V}_{22}-{\bf V}_{23}) \ , \nonumber \\
{\bf U}_{23}-{\bf U}_{22}&=&\cos\b({\bf V}_{23}-{\bf V}_{22})
  + \sin\b({\bf V}_{13}-{\bf V}_{12}) \ .
\end{eqnarray}
\noindent
If ${\bf V}_{12}={\bf V}_{13}$ is satisfied, with
  ${\bf V}_{22}\not={\bf V}_{23}$, the relation of eq.(11)
is reproduced. In general, this condition could exist for an orthogonal matrix.
However, we cannot get this solution without fixing the value of
$\tan\b$ in the mass matrix in eq.(7).\par
Studying the cases
${\bf O}_{11}={\bf O}_{13}$ and
${\bf O}_{11}={\bf O}_{12}$  does not add new conditions to
 our result, because the exchange of the rotation axis gives
the same conditions as in the above case.
Thus, we consider only the case in eq.(10)(or (12))
in order to get maximal $CP$ violation.\par
  Let us search for the constraints on the parameters in the Higgs potential
to give the orthogonal matrix {\bf U} in eq.(12).
These are easily found
by rotating the Higgs mass matrix ${\bf M^2}$  with
$\th=\b$ so as to make
 the (1,3)(and  then (3,1))
component  zero. The orthogonal matrix ${\bf U_0}$ is given by
\begin{equation}
 {\bf U_0}=\left(\matrix{\cos\b& \sin\b &0\cr -\sin\b& \cos\b &0\cr 0&0&1 }
   \right)\ .
\end{equation}
\noindent
The transformed matrix ${\bf M'^2=U_0^t M^2 U_0}$ is then given by
\begin{eqnarray}
M_{11}^{'2}&=&2g_1\cos^4\b+2g_2\sin^4\b+4(\bar\xi-g)\sin^2\b\cos^2\b\ ,
\nonumber\\
M_{22}^{'2}&=&2(g_1+g_2+2g-2\bar\xi)\sin^2\b \cos^2\b
+g'+\bar\xi+h\cos2\phi \ ,         \nonumber\\
M_{33}^{'2}&=& g'+ \bar\xi-h\cos 2\phi \ ,\nonumber\\
M_{12}^{'2}&=&\sin\b\cos\b \left[\cos 2\b(g_1+g_2+2g-2\bar\xi)+g_1-g_2
          \right ]\ ,  \\
M_{13}^{'2}&=&0 \ ,         \nonumber\\
M_{23}^{'2}&=&-h\sin2\phi \ ,\nonumber
\end{eqnarray}
\noindent
in $v^2\equiv |v_1|^2+|v_2|^2$ units and the parameter $\bar\xi$ is
                defined as   $\bar\xi=\xi/|v_1v_2|^2$.
We can easily find conditions to derive the orthogonal matrix ${\bf U}$ in
eq.(12), which do not depend on the specific values of
${\tan\b}$, as follows:
\begin{equation}
g_1+g_2+2g-2\bar\xi=0\ ,\qquad g_1=g_2\ ,\qquad \phi={\pi\o 4}\ .
\end{equation}
\noindent  The $CP$ violating phase $\phi$ takes its maximal value as is
expected.
The condition $g_1=g_2$ may be reasonable since
the infrared fixed point, approached using the renormalization group equations,
suggests $g_1\simeq g_2$[13].
The condition of $g_1+g_2+2g-2\bar\xi=0$ gives an important
constraint for the neutral Higgs scalars and the charged Higgs one,
since  ${\bar \xi}$ determines the charged Higgs mass as follows:
\begin{equation}
 m^2_{Hch}=\bar\xi v^2 \ .
\end{equation}
In addition to these constraints, we have the positivity condition
expressed as[13]:
\begin{equation}
g_1>0\ ,\quad g_2>0\ ,\quad h<0\ ,\quad h+g'<0\ ,\quad g+g'+h>-\sqrt{g_1g_2}\ .
\end{equation}

Under the condition of eq.(16), we have following
matrix elements for  ${\bf U}$:
\begin{eqnarray}
{\bf U}_{11}&=&\cos\b \ ,\qquad \qquad {\bf U}_{21}=-\sin\b \ ,\nonumber\\
{\bf U}_{12}&=&\cos\phi\sin\b\ , \qquad {\bf U}_{22}=
\cos\phi\cos\b   \ , \nonumber \\
{\bf U}_{13}&=&\sin\phi\sin\b\ ,\qquad {\bf U}_{23}=
\sin\phi\cos\b,
\end{eqnarray}
\noindent with $\phi=\pi/4$.
Then, the matrix elements of ${\bf O}$ are given by eq.(9) as:
\begin{equation}
{\bf O}_{11}=\cos2\b \ ,\quad {\bf O}_{12}=\sin\phi\sin2\b \ ,
\quad {\bf O}_{13}=\cos\phi\sin2\b \ .
\end{equation}
\noindent
We have two solutions yielding maximal $CP$ violation, which satisfy the
condition in eq.(6), for $\tan\b$:
\begin{equation}
\tan\b={1\o\sqrt{2}}(\sqrt{3}-1)=0.51\cdots \ ,\qquad
\tan\b={1\o\sqrt{2}}(\sqrt{3}+1)=1.93\cdots \ .
\end{equation}

 On the other hand, the masses of the three neutral Higgs scalars are given as
\begin{eqnarray}
m_{H1}^2&=&2g_1\cos^4\b+2g_2\sin^4\b+4(\bar\xi-g)\sin^2\b\cos^2\b
  =2g_1  \ ,\nonumber \\
m_{H2}^2&=& g'+ \bar\xi+h \ , \qquad
m_{H3}^2= g'+ \bar\xi-h \ ,
\end{eqnarray} \noindent in units of $v^2$, where
 the  conditions in eq.(16) are used in the second equality of $m_{H1}^2$.
We notice that the four Higgs masses
are given by four parameters $g_1$, $g'$, $h$ and $\bar\xi$.
Since the parameter $h$ is predicted to be very small in some
analyses[13],
 the values of $m_{H2}$ and $m_{H3}$ are expected to be almost degenerate.
The values of $m_{H2}$ and $m_{H3}$ are smaller than $m_{Hch}$ because
$g'+h$ is negative as seen in eq.(18).
On the other hand, $m_{H1}$  is not constrained in these discussions.  However,
   we use $700\G$ as the upper bounds of the Higgs masses given by the
perturbative anlyses of THDM[14].\par
 Let us study the $\rho$ parameter in the case of maximal $CP$ violation
with the Higgs masses in eq.(22).
 We set the SM with one Higgs doublet($m_{H{\rm ref}}$) as a reference
point and study the deviation from this point.
The extra contribution to $\rho$ in the $CP$ violating
THDM becomes[9,10]:
\begin{eqnarray}
& &\Delta\rho = {3\a\o 16\pi \cos^2\th_w} \sum_{i=1}^3
  {{\bf O}_{1i}^2\o m_Z^2-m_W^2} L(m_{Hi}^2, m_{H{\rm ref}}^2)
  \\
  &+& {\a\o 16\pi \sin^2\th_w m_W^2}\left[ \sum_{i=1}^3
   (1-{\bf O}_{1i})^2 F(m_{Hi}^2, m_{Hch}^2)-
  {1\o 2} \sum_{i,j,k=1\atop i\not=j, j\not=k, k\not=i}^3
  {\bf O}_{1i}^2 F(m_{Hj}^2, m_{Hk}^2)  \right], \nonumber
\end{eqnarray}
\noindent
where
 \begin{eqnarray}
 F(x,y)&=&{x+y\o 2}-{xy\o x-y}\log{x\o y}\ , \nonumber \\
 L(x,y)&=&F(x,m_Z^2)-F(x,m_W^2)+F(y,m_W^2)-F(y,m_Z^2) \ .
\end{eqnarray}

We present numerical results in the case of maximal $CP$ violation.
 The lower bound of the charged Higgs scalar mass has been
obtained by studying the inclusive decay $B\A X_s\r$[15],
as to which the upper bound of the branching ratio was recently
given by the CLEO collaboration[16]. The obtained
  lower bound is around $300\G$, which corresponds to $\bar\xi>3$.
Therefore, we take the charged Higgs scalar mass to be larger than $300\G$.
Then, the neutral Higgs scalar masses $m_{H2}$ and  $m_{H3}$ are taken to be
sma
   ller than $m_{Hch}$,
depending on the absolute values of  $g'$ and $h$
as seen in eq.(22).\par
We show the $m_{Hch}$ dependence on $\Delta\rho$ taking four  extreme parameter
   sets, (1)$g'=-0.5,\ h=-0.3,\ m_{H1}=100\G$,
(2)$g'=-2,\ h=-0.3,\ m_{H1}=100\G$, (3)$g'=-0.5,\ h=-0.3, m_{H1}=700\G$
and (4)$g'=-2,\ h=-0.3,\ m_{H1}=700\G$, in fig.1,
where the recent experimental value
$\rho=(3.0\pm 1.7)\times 10^{-3}$[17] is denoted by the horizontal dotted
lines.
Here, we take $m_{H{\rm ref}}=m_Z$ as a reference point.
The magnitude of $h$ is taken to be rather larger than usual[13]
in order to protect the under-estimate of $\Delta\rho$.
We consider two cases, that the value of $|g'|$ is small and that it is
large in order to find the $g'$ dependence of our predictions.
As seen in fig.1,  $\Delta\rho$ does not vary greatly relative to
the experimental value, and that our predictions are
 almost all negative except for the case (2)$g'=-2,\ h=-0.3,\ m_{H1}=100\G$,
in which we have a rather light neutral Higgs with a mass of $100\G$.\par
\begin{center}
\unitlength=0.7 cm
\begin{picture}(2.5,2.5)
\thicklines
\put(0,0){\framebox(3,1){\bf fig.1}}
\end{picture}
\end{center}

In order to find the $m_{H1}$ dependence of our result,
 we present $\Delta\rho$  versus $m_{H1}$ in four extreme cases,
 (5)$g'=-0.5,\ h=-0.3,\ m_{Hch}=350\G$, (6)$g'=-2,\
    h=-0.3,\  m_{Hch}=350\G$,
 (7)$g'=-0.5,\ h=-0.3,\ m_{Hch}=700\G$ and
(8)$g'=-2,\ h=-0.3,\ m_{Hch}=700\G$, in fig.2.
We found that our prediction is not large relative to
the experimental value unless  $m_{H1}$
is lower than $100\G$.
Thus, the THDM with maximal $CP$ violation is not contradicted with  the
present data of the $\rho$ parameter due to the conditions in eq.(16)
unless the extremely large value of $\mid g'\mid$ is taken,
even if the custodial symmetry is absent in the Higgs Lagrangian.
\begin{center}
\unitlength=0.7 cm
\begin{picture}(2.5,2.5)
\thicklines
\put(0,0){\framebox(3,1){\bf fig.2}}
\end{picture}
\end{center}

It may be useful to comment on the $CP$ violating parameter
 ${\rm Im}Z_i^{(k)}$, which is the imaginary part of
the $k$-th column vectors in the neutral Higgs scalar vector space,
defined in ref.11.
For the first Higgs scalar, these are zero because the third
component of the eigenvector is zero as seen in eq.(12),
i.e., there is no scalar-pseudoscalar interference term.
We have non-vanishing values for second and third Higgs scalars
(k=2,3) as follows:
\begin{equation}
  {\rm Im} Z_1^{(2)}=-{\rm Im} Z_1^{(3)}={1\o 4}(\sqrt{3}\mp 1)^2
 \ , \qquad
  {\rm Im} Z_2^{(2)}=-{\rm Im} Z_2^{(3)}=-{1\o 4}(\sqrt{3}\pm 1)^2
\ ,
\end{equation}
\noindent the signs $\pm$ correspond to the two solutions
 of $\tan\b$ in eq.(21).
We notice that these values are somewhat smaller than
   the Weinberg's bound[11] taking the same value of $\tan\b$ where
\begin{equation}
  \left |{{\rm Im} Z_1^{(2,3)}\o {\rm Im} Z_1^{(WB)}}\right | \simeq
\left\{ \matrix {0.89\cr 0.46} \right. \ , \qquad
 \left |{{\rm Im} Z_2^{(2,3)}\o {\rm Im} Z_2^{(WB)}}\right |\simeq
\left\{ \matrix {0.46\cr 0.89} \right.   \ ,
\end{equation}
\noindent
$(WB)$ denoting the Weinberg's bounds, and the upper values and lower
ones correspond to the two solutions of $\tan\b$. Thus, the Weinberg's
bound does not correspond to maximal $CP$ violation.
\par
We summarize as follows.
We have studied the conditions necessary to give maximal $CP$ violation
  in THDM with the $CP$ violation.
We obtained fixed values of $\tan\b$, and
 constraints on the Higgs potential parameters.
Two  neutral Higgs scalars must be lighter  than  the charged Higgs scalar
and these two Higgs scalars are expected to be
almost degenerate due to the smallness of the parameter $h$.
Under these conditions, the contribution of the Higgs scalar exchanges to
the $\rho$ parameter
 is rather small and negative in a wide range of Higgs masses.  Thus,
maximum $CP$ violation in THDM is safely
allowed for the $\rho$ parameter without the custodial symmetry.
\vskip 1 cm
\centerline{\bf Acknowledgments}\par
The authors would like to give thanks to Dr.B.E. Hanlon for reading the
manuscript.
One of us(M.T) thanks the particle group at Institut f\"ur Theoretisch
Physik in Universit\"at Wien, especially,
 Prof.H. Pietschmann, for kind hospitality.
 This research is supported  by the Grant-in-Aid for Scientific
Research,
Ministry of Education, Science and Culture, Japan(No.06220101 and No.06640386).
\newpage
\centerline{\large \bf References}
\vskip 1 cm
\noindent
[1] M.Kobayashi and T.Maskawa, Prog. Theor. Phys. {\bf 49}(1973) 652.\par
\noindent
[2] For a text of Higgs physics see J.F. Gunion, H.E. Haber, G.L.Kane
and S. Dawson,\par
  {\it "Higgs Hunter's Guide"},  Addison-Wesley, Reading, MA(1989). \par
\noindent
[3] G.C. Branco and M.N. Rebelo, Phys. Lett. {\bf 160B}(1985)117.
\par\noindent
[4] S. Weinberg, Phys. Rev. Lett. {\bf 63}(1989)2333;\par
  J.F. Gunion and D. Wyler, Phys. Letts. {\bf 248B}(1990)170;\par
  M. Chemtob, Phys. Rev. {\bf D45}(1992)1649;\par
     T. Hayashi, et al., Prog. Theor. Phys.(1994), in press.\par
\noindent
[5] S.M. Barr and A. Zee, Phys. Rev. Lett. {\bf 65}(1990)21;\par
  J.F. Gunion and R. Vega, Phys. Lett. {\bf 251B}(1990)157;\par
D. Chang, W-Y. Keung and T.C. Yuan,  Phys. Rev. {\bf D43}(1991)14;\par
R.G. Leigh et al., Nucl. Phys. {\bf B352}(1991)45.\par
\noindent
[6] B. Grz\c adkowski and J.F. Gunion, Phys. Lett.
                            {\bf 287B}(1992)237;\par
 B. Grz\c adkowski and W-Y. Keung,
    preprint at CERN, CERN-TH.6941/93(1993);\par
 C. Schmidt and M. Peskin, Phys. Rev. Lett. {\bf 69}(1992)410;\par
R. Cruz, B. Grz\c adkowski and J.F. Gunion, Phys. Lett.
                            {\bf 289B}(1992)440;\par
D. Atwood, G. Eilam and A. Soni, Phys. Rev. Lett.
       {\bf 70}(1993)1364;\par
 B. Grz\c adkowski , preprint at
        Warsaw University, IFT07/94(1994).\par
\noindent
[7] A. M\'endez and A. Pomarol, Phys. Lett.
                            {\bf 272B}(1991)313.\par
\noindent
[8] B. Grz\c adkowski and J.F. Gunion, Phys. Lett.
                            {\bf 294B}(1992)361.\par
\noindent
[9] C.D. Froggatt, R.G. Moorhouse  and I.G. Knowles,
   Phys. Rev. {\bf D45}(1992)2471;\par
   Nucl. Phys. {\bf B386}(1992)63.\par
\noindent
[10] A. Pomarol and R. Vega, preprint at SLAC, SLAC-PUB-6169(1993).\par
\noindent
[11] S. Weinberg, Phys. Rev. {\bf D42}(1990)860.\par
\noindent
[12]  H. Georgi, Hadr. J. Phys. {\bf 1}(1978)155.\par
\noindent
[13] M.A. Luty, Phys. Rev. {\bf D41}(1990)2893;\par
     C.D. Froggatt, I.G. Knowles, R.G. Moorhouse,
     Phys. Lett. {\bf 249B}(1990)273; \par
  M. Chemtob, Z. Phys. {\bf C60}(1993)443.\par
 \noindent
[14] D. Kominis and R.S. Chivukula,
      Phys. Lett. {\bf 304B}(1993)152. \par
 \noindent
[15] T.Hayashi, M.Matsuda and M.Tanimoto,
          Prog. Theor. Phys. {\bf 89}(1993)1047.\par
 \noindent
[16]  R. Ammar et al., CLEO Collaboration,
     Phys. Rev. Lett. {\bf 71}(1993)674.\par
     \noindent
[17] R. Barbieri, preprint at Pisa University,
                             IFUP-TH 28/94(1994).\par

\newpage
\topskip 1 cm
\centerline{\large \bf Figure Captions}\par
\vskip 1 cm
\noindent
{\bf Fig.1}: The contributions of the Higgs scalar
    exchanges to the $\rho$ parameter
   versus the charged Higgs scalar mass.
 The cases (1), (2), (3) and (4) are denoted by
 the solid, dotted, dashed, dashed-dotted curves, respectively.
 The horizontal dashed lines denote the experimental bounds.
\vskip 2 cm
\noindent
 {\bf Fig.2}: The contributions of the Higgs scalar
   exchanges to the $\rho$ parameter
    versus  $m_{H1}$.
 The cases (5), (6), (7) and (8) are denoted by
 the solid, dotted, dashed, dashed-dotted curves, respectively.
\end{document}